\documentclass[prd,aps,epsfig,floats,superscriptaddress,showpacs]{revtex4}  
\usepackage{graphics} 
\tighten

\newcommand{\be}{\begin{equation}} 
\newcommand{\ee}{\end{equation}} 
\newcommand{\bea}{\begin{eqnarray}} 
\newcommand{\eea}{\end{eqnarray}}  
\newcommand{\bean}{\begin{eqnarray*}} 
\newcommand{\eean}{\end{eqnarray*}}

\def\lsim{\raise 0.4ex\hbox{$<$}\kern -0.8em\lower 0.62ex\hbox{$\sim$}} 
\def\gsim{\raise 0.4ex\hbox{$>$}\kern -0.7em\lower 0.62ex\hbox{$\sim$}}

\begin{document} 
%\draft 
%\twocolumn[\hsize\textwidth\columnwidth\hsize\csname@twocolumnfalse\endcsname
\title{Generation of Primordial    
Cosmological Perturbations from Statistical Mechanical Models}   
\author{A. Gabrielli}   
\affiliation{INFM UdR Roma1,       
Dip. di Fisica, Universit\'a ``La Sapienza'',     
P.le A. Moro, 2, I-00185 Roma, Italy. }    
\author{B. Jancovici}    
\affiliation{Laboratoire de Physique Th\'eorique,    
         Universit\'e de Paris XI, B\^atiment 211,   
  91403 Orsay, France \footnote{Unit\'e Mixte de Recherche no. 8627-CNRS}.} 
\author{M. Joyce}   
\affiliation{Laboratoire de Physique Th\'eorique,   
         Universit\'e de Paris XI, B\^atiment 211,   
  91403 Orsay, France \footnote{Unit\'e Mixte de Recherche no. 8627-CNRS}.}  
\affiliation{Laboratoire de Physique Nucl\'eaire et de Hautes Energies,  
 Universit\'e de Paris VI, 4, Place Jussieu, 
Tour 33 -Rez de chaus\'ee, 75252 PARIS Cedex 05 \footnote{Permanent   
address from Sept. 1st 2002.}.}  
\author{J. L. Lebowitz}    
\affiliation{Department of Mathematics, Rutgers University, New  
Brunswick, NJ08903 USA.}   
\author{L. Pietronero}    
\affiliation{INFM Sezione Roma1,        
Dip. di Fisica, Universit\'a ``La Sapienza'',     
P.le A. Moro, 2, I-00185 Roma, Italy. }    
\author{F. Sylos Labini}  
\affiliation{Laboratoire de Physique Th\'eorique,    
         Universit\'e de Paris XI, B\^atiment 211,   
  91403 Orsay, France \footnote{Unit\'e Mixte de Recherche no. 8627-CNRS}.}  
\affiliation{INFM Sezione Roma1,       
Dip. di Fisica, Universit\'a ``La Sapienza'',     
P.le A. Moro, 2, I-00185 Roma, Italy. }  
\begin{abstract}   
\begin{center}    
{\large\bf Abstract}   
\end{center}    
The initial conditions describing seed fluctuations for the formation of    
structure in standard cosmological models, i.e.\ the Harrison-Zeldovich    
distribution, have very characteristic ``super-homogeneous'' properties: 
they    
are statistically translation invariant, isotropic, and  
the variance of the mass fluctuations in a region of volume $V$   
grows slower  
than $V$.  We    
discuss the geometrical construction of distributions of points in   
${\bf R}^3$ with similar properties encountered in tiling and in    
statistical physics, e.g. 
the Gibbs distribution of a one-component system of charged   
particles in a uniform background (OCP).  Modifications of the OCP can  
produce   
equilibrium  correlations of the kind assumed in the cosmological  
context.  We then describe how such systems can be used for the   
generation of initial conditions in gravitational $N$-body simulations.  
\end{abstract}    
\maketitle   
\date{today}  
%]    
\twocolumngrid   
     
\section{Introduction}   
    
A central problem in contemporary cosmology is the quantitative explanation  
of the inhomogeneity observed in the Universe at large scales (see    
e.g. \cite{pee93,padm}).  Such inhomogeneity is probed both indirectly for  
early times via    
the fluctuations in temperature in the microwave sky, and directly at  
present in the  
distribution of matter in space.
All currently standard cosmological models    
work within a paradigm in which these fluctuations are the result of    
(mostly) gravitational evolution operating on some initial very small    
fluctuations.  Common to all such models is the assumption - supported in    
particular by observations of the microwave background \cite{cobe}- of a   
very specific form of these initial perturbations at large scale, i.e.\  
Gaussian fluctuations with a Harrison-Zeldovich (HZ, often referred to as 
``scale-invariant'')   power  
spectrum \cite{har,zel}.  In a recent paper some of us \cite{hz} have   
discussed how, in a simple classification of correlated systems, this   
spectrum corresponds to highly ordered glass-like or lattice-like   
distributions. In particular such systems are characterised in real  
space by the striking behaviour of the variance in the mass, $\langle 
\Delta M^2 \rangle$, contained in a 
given volume, e.g.\ in spheres of radius $R$,  
$\langle \Delta M^2 \rangle \propto R^2$, or $R^2 \log R$, i.e.\ it is 
essentially    
proportional to the {\it surface area} of the sphere. 
This is to be contrasted with the Poisson behaviour   
(proportional to the volume $R^3$) characteristic
of many equilibrium systems with   
short range interactions away from phase transitions  
or that characteristic of long-range correlations    
at a critical point in many systems (for which 
$\langle \Delta M^2 \rangle    
\propto R^\alpha$ with $\alpha>3$).   
Actually $R^2$ is the   
slowest growth possible for {\it any} isotropic    
translationally invariant distribution of points \cite{beck}.   

In this paper we examine the nature of the correlations in such   
super-homogeneous systems and compare them with    
equilibrium correlations in systems with long range interactions  
studied in statistical   
mechanics 
\cite{martin-yalcin,levesque-weis-lebo,review-martin}. 
More precisely we consider the  
One Component Plasma (OCP), which is simply a system of    
charged point particles interacting through a repulsive 
$1/r$ potential, in a    
uniform background which gives overall charge 
neutrality (for a review, see    
\cite{OCP-review}).  We discuss how the OCP can be modified 
so that its equilibrium correlations are precisely those considered in    
cosmology. This involves an appropriately designed attractive short range  
potential as well as a repulsive $1/r^2$ potential at large    
scales. The latter corresponds to a four dimensional Coulomb potential  
with the particles confined to three dimensions.    
    
Our analysis of these parallels has more than    
theoretical interest. It also provides a new method for    
generating initial conditions (IC) for numerical studies     
of the formation of structure in cosmology \cite{HE,white93}.    
We explain that    
this method avoids the problems associated with currently    
used algorithms which involve superimposing small perturbations on a   
lattice or glass-like distribution. The latter    
is understood to be a ``sufficiently uniform'' discretization    
of a constant density background, not as representing the    
perturbed distribution. The problem with this choice of initial state  
is that the {\it unperturbed} distribution has    
its own inherent fluctuations/correlations, which can play an
important role in the evolution of the system
\cite{melott,bsl02,bjsl02,slbj02}.

%%%%%%%%%%%%%%%%%%%%%%%%%%%%%%%%%%%%%%%%%%%%%%%%%%%%%%%%%%%%%%%%%%%   

\section{The HZ spectrum in cosmology}    
   
Let us recall first the necessary essentials of modeling perturbations from 
a uniform density used in   
cosmology. These are described as a {\it continuous} stationary   
stochastic process (SSP),   
% The stochasticity refers to the fact that we   
%suppose our universe one of an ensemble described by some probability   
%density functional,  
with the property  
%of {\it ergodicity} guaranteeing  
that the volume average of observables, i.e.\ empirical    
averages,  approach the   
ensemble averages in the large volume limit. The stationarity refers in   
this context to the statistical invariance under spatial translation.   
Moreover one assumes also statistical isotropy of the  stochastic   
process, i.e. invariance under rotation of the ensemble   
quantities. Thus if $\rho(\vec{r})$ is the density field, we have   
\be      
\label{shi2a}      
\left< \rho(\vec{r}) \right> = \rho_0 > 0\,,       
\ee    
and we can define a {\it reduced} 2-point correlation function   
$\tilde\xi(r)$ by   
\be         
\label{shi4}        
\left<\rho(\vec{r}_{1}) \rho(\vec{r}_{2})\right> \equiv       
\rho_0^2\left[1+\tilde\xi(r)\right]    
\ee    
where $r=|\vec{r}_{1} - \vec{r}_{2}|$. Corresponding isotropies are assumed
to hold for    
the higher order correlation functions.    
Further the probability distribution for the fluctuations in the 
initial uniform density are generally assumed to be   
Gaussian.  

Much more used in cosmology than $\tilde\xi(r)$ is the   
equivalent $k$-space quantity, the power spectrum $P(\vec{k})$ which    
is defined as    
\be   
P(\vec{k})=\lim_{V\rightarrow\infty}
\frac{\left<|\delta_\rho(\vec{k})|^2\right>}{V}    
\label{ps-defn}    
\ee     
where   
$\delta_\rho(\vec{k})=\int_V d^3r e^{-i\vec{k}\cdot \vec{r}} \delta (\vec{r})$
is the Fourier integral in the volume   
$V$ of the normalized fluctuation field 
$\delta(\vec{r})=(\rho(\vec{r})-\rho_0)/\rho_0$.   
In a statistically isotropic SSP this depends only on $|\vec{k}|$ with   
\be   
P(\vec{k}) \equiv P(k) = \int_0^{\infty}    
\tilde\xi({r})  \frac{\sin (kr)}{kr} 4 \pi r^2 dr \,.    
\label{lat7a}      
\ee      
It follows from its definition that   
$P(k)\ge 0$, and, in the case we are considering, 
its integral, which is equal to $\tilde\xi(0)$, is assumed   
to be finite.   

In current cosmological models it is generally assumed that the   
power spectrum $P(k) \propto k$ at small $k$. This is known as the HZ   
or ``scale-invariant'' power spectrum. It is believed to describe the     
``primordial'' fluctuations at very early times, the putative 
remnants of a period of ``inflation'' prior to the ordinary
Big Bang phase \cite{pee93,padm}.    
The linear $k$  behaviour is cut-off by a short distance   
scale (needed to ensure integrability) i.e. for $k$ larger   
than some cut-off, $k_c$, $P(k)$ decreases faster than
$k^{-3}$. The reason for the appearance of this particular 
spectrum is tied to considerations   
about the  cosmological model (see \cite{pee93, padm}).   
Thus in the homogeneous and isotropic 
Friedmann-Robertson-Walker (FRW) cosmology, the only characteristic length 
scale is the size of the horizon $R_H(t)$ (the region
causally connected at a time $t$).  The HZ spectrum
then corresponds to the choice which gives 
(as discussed in \cite{hz} this condition   
is in fact only satisfied for a spherical Gaussian
window function, and not in a real physical sphere) 
\be    
\sigma_M^2 (R_H(t)) = {\rm constant},
\label{HZ-criterion}
\ee    
independent of $t$,where $\sigma_M^2 (R)$ is the normalized mass variance
\be 
\label{variance-def} 
\sigma_M^2 (R)=\left<\Delta M^2\right>/\left<M\right>^2 \,. 
\ee   
The HZ spectrum then follows from a
consistency criterion for the treatment of the
perturbations in the model: any other spectrum will give
mass fluctuations which dominate over the homogeneous
background either at some time in the future or past.
   
All current cosmological models share the HZ spectrum (or something
very close to it) as the ``primordial'' form of
their perturbations. This
spectrum evolves, at larger and larger scales, until
the characteristic time at which the densities of
matter and radiation energy are equal.  This evolution which depends
strongly on the particular model used may change completely
the form of the spectrum for distances smaller than  a characteristic
scale $k_{eq}^{-1}$ (i.e. $k > k_{eq}$).
For larger physical scales ($k < k_{eq}$), however,
the evolution is almost exactly the same in all models, and
it leaves the primordial HZ form intact.

Viewed in the general framework of correlated processes  
a very crucial property of the HZ type power spectrum is 
that  
\be
\label{limitk=0}
\lim_{k\rightarrow 0} P(k)=0
\ee 
which in real space implies that 
\be    
\lim_{R\rightarrow \infty} {\langle \Delta M^2 \rangle \over V(R)} = \lim_{R\rightarrow \infty} \int_{V(R)}  d^3r \; \tilde \xi(r) = 0     
\label{integral-constraint} 
\ee
where $V(R)$ is the volume of a sphere of radius $R$ (with arbitrary
origin). Eq. (\ref{integral-constraint}) is to be contrasted with 
the behaviour of this integral in a Poisson distribution, when
it yields a finite positive constant,
and with that associated with power-law correlations
in critical systems \cite{hz} for which the same integral
of $\tilde{\xi}(r)$ diverges.
   
Systems satisfying Eq. (\ref{integral-constraint}) 
are thus ``more homogeneous'' than a Poisson type system,
as can be seen when one considers the 
behaviour of the mass variance in spheres. For a spectrum    
such that $P(k) \sim k^n$ for $k \rightarrow 0$
and appropriately cut-off at large $k$ we have for large
$R$   
\be   
\sigma_M^2(R) \propto   
\left\{ \begin{array}{lll} 
 1/R^{3+n} \; \;  \mbox{if} \;\; n<1\\   
\log(R)/R^{4} \;\; \mbox{if} \;\;  {n=1}  \\  
1/R^{4}  \;\; \mbox{if} \;\;  {n>1} 
\end{array}  
\right. 
\ee  
In terms of the non-normalized
quantity $\left<(\Delta M)^2\right> \propto \sigma_M^2(R)R^6 $,
Eq. (8) says that for $n > 0$
we have a slower increase of
the variance as a function of $R$ than for Poisson fluctuations,
corresponding to $n=0$, with the limiting behaviour 
corresponding to a variance which is proportional
to the {\it surface area} of the sphere. These systems are 
thus characterised by surface fluctuations,    
ordered (or homogeneous enough, one could say) to 
give this very particular behaviour. In Fig. 1
are shown, in two dimensions, a Poissonian distribution 
and a super-homogeneous distribution. The relatively greater
uniformity of the latter is clearly identifiable.
\begin{figure}[tbp]
\scalebox{0.8}{\includegraphics{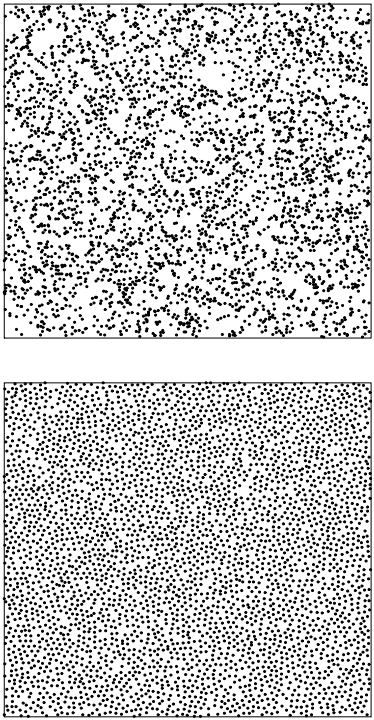}}     
\caption{A super-homogeneous distribution (bottom) and a Poisson 
distribution (top) with approximately the same number of points. 
Both are projections of thin slices of three dimensional distributions. 
The super-homogeneous one is a ``pre-initial'' configuration used 
in setting up the configuration representing the initial
conditions for the cosmological N-body simulations of \cite{virgo}. 
As described in section \ref{sectionNBS} 
the dynamics used to generate this ``pre-initial''configuration 
gives rise to surface fluctuations as it is essentially
the one component plasma which we discuss below. 
\label{fig1}}
\end{figure}
 
\section{Distributions of points with surface growth of the squared variance}
A simple cubic lattice is the simplest example of a discrete
set of points which shows this limiting behaviour
($\sigma_M^2 \propto 1/R^4$) of the variance \cite{kendall}. 
The result is not hard to understand: The standard deviation $\sigma_M$ is  
that measured by averaging over spheres of radius $R$ centered at a
randomly chosen point of the unit cell. It is proportional to the
typical variation of points in the volume, that in the case of a
lattice is the square root of the average number of points in the last
spherical shell of thickness equal to the lattice unit. A better (more
statistically uniform) example of the same kind is the 
so-called {\em shuffled lattice} \cite{hz}; 
this is a lattice whose sites are independently randomly displaced 
by a distance $x$ in all directions from their initial position according to
some distribution $p(x)$ which has a finite second moment. In
this case we find $P(k)\sim k^2$ at small $k$ and, consequently, again
$\sigma_M^2(R)\propto 1/R^4$ at large $R$. The simple lattice, however,
is not a SSP, and even the shuffled lattice, though it can be defined 
as a SSP, is not statistically isotropic because the underlying 
lattice structure is not completely erased by the shuffling \cite{gacs}.  

\begin{figure}[tbp]      
\scalebox{0.5}{\includegraphics{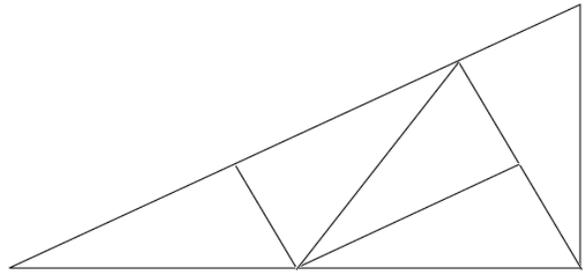}}     
\caption{Fragmentation step for the ``pinwheel'' tiling of  
the plane.)   
\label{fig2}  } 
\end{figure}

To construct a statistically isotropic and homogeneous particle 
distribution with such a behaviour of $\sigma^2(R)$ is
non-trivial.  A particular example is the so-called  
``pinwheel'' tiling of the plane \cite{radin1,radin2}. The 
generation algorithm for it is 
defined by taking a right angled triangle with sides of respective 
length one and two (and hypotenuse $\sqrt 5$) and, at the first step, 
forming five similar square triangle of sides $1/\sqrt 5$ and $2/ \sqrt 5$
respectively as shown in Fig.~\ref{fig2}. At the second step we 
expand these new triangles, to the size of the original triangle, and
repeat the procedure {\em ad infinitum}, so that they cover the plane 
completely.  
Finally, placing a point randomly inside each elementary triangle  
will give the superhomogeneous point distribution which is  
statistically isotropic (with a continuous power spectrum). 
   
\section{A physical example: The One-Component Plasma}   
   
In the study of the one component plasma (OCP) one considers the
equilibrium statistical mechanics (at inverse temperature $\beta$) of
a system of charged point particles interacting through a Coulomb
potential, in a continuous uniform background giving overall charge 
neutrality \cite{OCP-review}. Taking the particles to carry
unit mass and charge the
microscopic mass density of the particles is given by
\be
\rho(\vec{r})= \sum_i \delta(\vec{r}-\vec{r}_i)
\label{micro-density-particles}
\ee 
where $\vec r_i$ is the position of the $i$th particle.  The
total microscopic charge density is 
\be 
\rho_c(\vec{r})= \sum_i \delta(\vec{r}-\vec{r}_i) -n 
\label{micro-density}
\ee   
where   
\be   
n=\langle\rho(\vec{r})\rangle \;.
\label{mean-density}
\ee 
The two-point correlation function is defined as
\bea   
\tilde{\xi} (\vec{r}-\vec{r}\,')&=&    
 \frac{\langle \rho(\vec{r}) \rho(\vec{r}\,') \rangle}    
{\langle \rho \rangle^2} -1    
\nonumber \\    
&=&    
\frac{\delta(\vec{r}-\vec{r}\,')}{n} + h (\vec{r}-\vec{r}\,')\,.    
\label{reduced-density-density}   
\eea     
The first term comes from the diagonal terms $\vec{r}=\vec{r}\,'$ 
in $\langle \rho(\vec{r}) \rho(\vec{r}\,') \rangle$    
and is characteristic of any point distribution. 
    
It can be proven quite generally \cite{review-martin}
that at high temperatures the OCP has a translation invariant isotropic 
distribution with $\tilde{\xi} (\vec{r})$ satisfying 
Eq. (\ref{integral-constraint}) with $r = |\vec{r}|$.
In terms of $h(r)$ this gives 
\be 
n \int h(r) d^3 \vec{r} = -1   
\label{integral-h}  
\ee 
Eq.(\ref{integral-h}) is usually referred to as a sum rule  
which originates from the perfect screening of each charge  
caused by the long-range nature of the Coulomb potential. 
To see this let us suppose 
that an external infinitesimal charge density
$\rho_{\rm{ext}}=\epsilon e^{i\vec{k}.\vec{r}}$ of very long
wavelength is applied to the system. It creates an
external electric potential    
\be 
\Phi(\vec{r})= 
\int \frac{\rho_{\rm{ext}}(\vec{r}'\,)}{|\vec{r} -\vec{r}\,'|} d^3 \vec{r}\,'  
=\frac{4\pi}{k^2} \epsilon e^{i\vec{k}.\vec{r}} 
\label{external-potl}
\ee 
and a perturbation to the Hamiltonian  
\be 
H_{\rm{ext}}
=\int \rho_c(\vec{r}) \Phi(\vec{r}) d^3 \vec{r}
=\epsilon \frac{4\pi}{k^2} \int \rho_c(\vec{r})
e^{i\vec{k}.\vec{r}} d^3 \vec{r} \;.
\label{external-hamiltonian}
\ee
Using linear response theory the induced charge in the
system is given by
\be  
\rho_{\rm{ind}}(\vec{r}\,') 
=-\beta \langle \rho_c(\vec{r}\,')  H_{\rm{ext}} \rangle 
\label{induced-charge}
\ee 
where the average is over the unperturbed statistical distribution.
Thus, assuming that the applied charge is perfectly screened  
i.e. the system responds with an induced charge density     
$\rho_{\rm{ind}}=-\rho_{\rm{ext}}$ we have, in the limit,
$k \rightarrow 0$, 
\be 
-\epsilon e^{i\vec{k}.\vec{r}\,'} \sim 
-\beta \epsilon \frac{4\pi}{k^2} \int
\langle \rho_c(\vec{r}\,')  \rho_c(\vec{r}) \rangle 
e^{i\vec{k}.\vec{r}} d^3 \vec{r} 
\label{induced}
\ee
and therefore, since
\be 
P(k) = \frac{1}{n} +  \tilde{h}(k)
\label{ps} 
\ee
where $\tilde{h}(k)$ is the Fourier transform of $h$, 
we find that
\be 
P(k) \sim \frac{k^2}{4\pi n^2 \beta}
\label{S-smallk}
\ee
for small $k$. 

The behaviour of the power spectrum at small $k$ is    
traceable through Eq. (\ref{external-potl}) as being simply,
up to a factor $\beta$,
the inverse of the Fourier transform,
(in the sense of distributions) of the repulsive $1/r$ potential.
It is evident that what one needs to  
obtain the same kind of behaviour but with $P(k) \sim k$ at small $k$ 
is a repulsive potential of which the Fourier transform 
behaves as $1/k$, i.e.\ a repulsive $1/r^2$ potential. 
Before considering this in more detail we discuss some issues 
relevant to the cosmological context which we have passed over
without comment so far. 
   
\section{Discrete vs. Continuous}  

An essential difference between the case of cosmological perturbations
and the system just discussed is that the former refers to a continuous
density field while the latter describes the correlation properties of
a set of discrete points. To interpret the latter as giving a realization
of the former we obviously need to specify explicitly how to relate the  
two. There is no unique prescription to pass from a discrete to a
continuous field. A prescription simply corresponds to a
regularisation of the Dirac delta with a function $ W_L(\vec{r})$
with the property
\bea  
W_L(\vec{r})=L^{-3} W_o(\frac{\vec{r}}{L})\,,
\qquad \int W_o(\vec{r}) d^3 \vec{r}=1  
\label{regularise}   
\eea   
where $L$ is the characteristic scale introduced by the 
regularisation e.g. the Gaussian 
\be 
W_L(\vec{r})=\left(\frac{1}{\sqrt{2\pi}L}\right)^{3}
\exp \left(-\frac{r^2}{2L^2}\right) \,. 
\label{regularise-gaussian}
\ee
For any finite value of $L$ we can define a continuous density field 
$\rho_L (\vec{r})$ as the convolution of this function with the
discrete density field. The pair correlation function of the continuous 
field can then be written also as a double convolution integral of 
$W_L(\vec{r})$ and the correlation function in the discrete case. 
The singularity in the latter is thus also removed by the 
regularisation. The power spectrum of
this continuous field is then simply given as
\be  
P_L(\vec{k})=|W_L(\vec{k})|^2 P_D(\vec{k})  
\label{regularise-ps}
\ee
where $W_L(\vec{k})$ is the Fourier transform of the regularisation
$W_L(\vec{r})$ and $P_D(\vec{k})$ is the power spectrum of the discrete
field. In particular for
the Gaussian smoothing function  we have
\be
P_L(\vec{k})= \exp (-k^2 L^2) P_D(\vec{k}) \,.
\label{regularise-ps-gauss}
\ee  

For the OCP discussed above the correlation function, in the
weak coupling limit, is approximately the Debye-H\"uckel
formula
\be 
h(r) = -\beta \frac{e^{-\kappa r}}{r} \qquad \kappa^2=4\pi\beta n
\label{high temperature}
\ee
so that
\be 
P_D(\vec{k}) =  \frac{1}{n} \left[\frac{k^2}{\kappa^2+k^2} \right]
\label{OCP-weakcoupling-ps}
\ee
which is a non-integrable function of $k$.  
Integrability  
is ensured by the cut-off imposed at large $k$ in the power
spectrum through the smoothing function.  
An assumed property of the primordial density field in standard   
cosmological models is that it is Gaussian distributed. Since the
density field is inherently positive this assumption of Gaussianity 
is more properly attributed to fluctuations (expected to be small) 
around the uniform density.
Small fluctuations in the discrete OCP are in fact Gaussian 
\cite{martin-yalcin}.

\section{Initial Conditions in N-body simulations}  
\label{sectionNBS}
   
A context in which it is necessary to have a concrete realization 
of the density $\rho(\vec{r})$ in cosmological models is that of
the N-body 
approach to the problem of structure formation \cite{HE, white93}.  
In these numerical 
simulations an initial configuration (IC), which should represent 
the universe at early times, is evolved under its self-gravity.  The goal
of these simulations is an understanding of how structures grow in 
such models, and whether they are compatible with observations of the 
distribution of matter (as probed, primarily, by the distribution of 
galaxies and galaxy clusters). The initial continuous density field 
is thus necessarily represented by a discrete set of points, which, 
because the discretization scale does not represent the real underlying 
physical one, must be considered just as we have discussed above as 
representing the continuous density field via some appropriate smoothing.   

In practice such a discretization is always generated in 
a very particular way \cite{EDFW, white93}: points of equal mass are  
displaced from a ``pre-initial'' configuration in a 
way prescribed by the correlation field one
wants to set up. 
This prescription can be understood as follows. 
Superimposing an infinitesimal displacement field
$\vec{u}(\vec{r})$ on a uniform density $\rho_o$ one has
\be   
\frac{\rho(\vec{r})-\rho_o}{\rho_o} \simeq - \vec{\nabla}.\vec{u}(\vec{r})  
\label{displacement} 
\ee   
and therefore in $k$-space the power spectrum of the correlated
density field $\rho(\vec{r})$ produced
is roughly proportional to $k^2 P_u(k)$,
where $P_u(k)$ is 
that of the displacement field. 
The ``pre-initial'' configuration used, which gives 
the initial unperturbed positions of the particles, is
a discretization of the initial uniform density field $\rho_o$
and the displacement field is specified by the power spectrum
of the continuous field which is desired as IC 
in the simulation. In practice the former is taken either to be
an exact simple lattice or a ``glassy'' configuration \cite{white93}.
When this latter configuration is used 
it is generated by evolving the system  
with the N-body code, but with the sign of gravity reversed. This
corresponds to a system essentially like the OCP (at low 
temperature), and the
long-time evolution brings it to a highly uniform configuration
with normalized variance decaying at large scales as $1/R^4$
\cite{bsl02}.     
   
In the generation of the IC in this way the   
fluctuations associated with the ``pre-initial'' configuration are   
neglected, or rather implicitly assumed to be negligible with respect 
to those which are introduced by the perturbations to it.  The problem 
with this procedure is that it neglects precisely 
these ``pre-initial'' fluctuations. 
(If we assume the initial continuous field
to have power-spectrum $P_i(k)$ what results 
from the infinitesimal shuffling is a power spectrum
$P(k)=P_i(k) + k^2 P_u(k)$). 
Such a neglect is justified if
these fluctuations play no role in the evolution of the 
system, as one  
wants to see only the growth of structure coming from 
the perturbations superimposed on the uniform density field. 
Indeed the reason the
lattice or ``glassy'' configuration is used is that they 
are 
configurations in which the gravitational force is (effectively) zero,
whereas in a Poisson distribution for example (which is a priori as
good a discretization of the uniform background) it is not. While the
Poisson distribution will thus evolve and form structures even
without additional perturbation, the
perfect lattice will not. That the discretized lattice or glass solves
this problem is however far from clear: they both
represent 
unstable point configurations with respect to gravity. A small applied
perturbation will thus in general give an evolution which depends on
the correlations and/or fluctuations in the initial unperturbed
configurations. In principle such effects can be kept under control by
making the discretization of the original density sufficiently 
fine. In practice in N-body simulations the ``non-linear regime'' in
which structure formation takes place is comparable to the
discretization scale, and the role of such effects seems, at the very
least, problematic 
\cite{melott,bsl02,bjsl02,slbj02}.     
  
One way in which this difficulty could be got around is by 
generating directly a discrete distribution whose regularisation  
is exactly the desired continuous field. Our observations on 
the OCP provides in principle a way of realizing this possibility.

\section{IC as equilibrium of a modified OCP}
  
We have seen that the OCP equilibrium correlations
give surface fluctuations 
($\langle \Delta M^2 \rangle \sim  R^2$), but
with a power spectrum at small $k$ which goes
like $k^2$.  
By considering instead a repulsive $1/r^2$ potential, whose
Fourier transform 
is $2\pi^2/k$, we obtain from Eq. (\ref{ps})
\be  
P(k) \sim \frac{k}{2\pi^2 n^2 \beta},\, {\rm for}\,  k \sim 0 
\label{S-smallk-1/r2}
\ee   
and 
\be  
h(r) \sim -\frac{1}{2\pi^4 \beta n^2}\frac{1}{r^4}
\label{h-smallk-1/r2}
\ee
for $r \gg (2\pi^2 \beta m)^{-1}$. The change from the exponential 
decay of Eq. (\ref{high temperature}) to a power-law decay
is a result of the different analyticity properties of the
two power-spectra: the $k^2$ behaviour is analytic at the 
origin, guaranteeing a rapidly decaying behaviour of its   
Fourier transform, while the $k$ spectrum is not. 
 
\begin{figure}[tbp]     
\scalebox{0.5}{\includegraphics{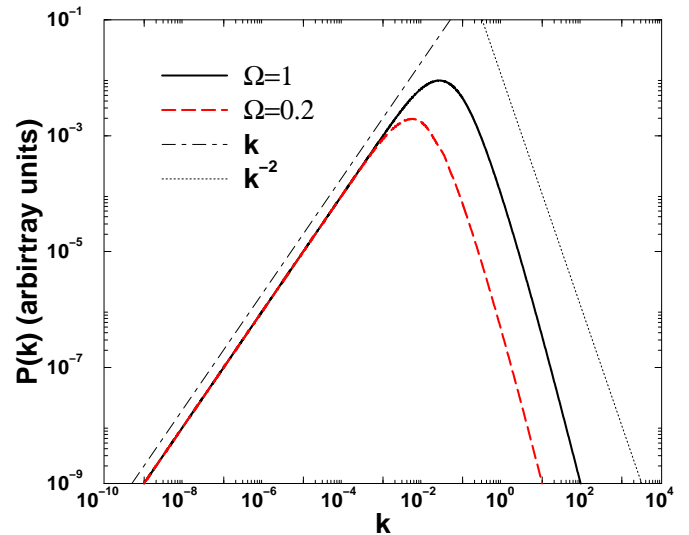}} 
\caption{Power spectra of two different cosmological 
models, both of the CDM (``cold dark matter'') type  
but with a differing density of matter (parametrised by $\Omega$).  
The plot is in log-log units so that the behaviour 
$P(k) \sim k$ and $P(k) \sim k^{-2}$ correspond to 
straight lines of slope $+1$ and $-2$ respectively.
\label{fig3}  } 
\end{figure}

In the context of cosmological N-body simulations what one
needs is not simply the primordial HZ power spectrum with
some appropriate small scale cut-off: what is simulated   
is only a part of the cosmological evolution, 
starting from a time at which the initial spectrum of
fluctuations is already significantly modified from
its primordial form. While purely gravitational
evolution 
at these early times does not modify the HZ spectrum,
non-gravitational effects, present until the time when the universe
becomes dominated by matter, do so significantly. The nature of 
these modifications depends on the details of the cosmological model,
but in all cases it affects only $k$ larger than a
characteristic $k_{eq}$ corresponding to the causal horizon
at the ``time of equality'' (when matter starts to dominate over
radiation). One then has a power spectrum of the form
\be
P(k) = k^n f(k)
\label{initial-ps}
\ee   
with $n$ exactly (or very close to) unity and $f(k)$ a model- 
dependent form for the spectrum with $f(0)=a>0$. In Figure \ref{fig3}
are shown the power spectra of two representative cosmological
models \cite{pee93, padm}. Both
are CDM (``cold dark matter'') type. HDM (``hot dark 
matter'') models, now strongly disfavored, have a more
abrupt (typically exponential) cut-off at larger $k$, due 
to the ``wash-out'' of small scale structure associated with
their higher velocity dispersions. 
   
To produce such a spectrum as the equilibrium one of an OCP like system   
requires further modification of the form of the potential.  
Just like the standard OCP, an unmodified $1/r^2$ potential will give, 
in the weak-coupling limit, a spectrum which becomes flat 
(i.e. Poisson like) at large $k$.  This is simply due to the
fact that in this limit the thermal fluctuations dominate 
over the potential at small scales, effectively
randomizing the particle positions up to some scale. 
A crude guess of what potential would produce the
behaviour of a typical cosmological model can be 
obtained by supposing  
that at small scales  
\be   
1+h(r) \approx e^{-\beta V(r)}   
\label{shortdistance-cfn}   
\ee   
which corresponds to completely neglecting collective effects   
($1+h$ represents the relative probability compared to random   
of finding a particle at distance $r$ from a given one).   
Given that the desired fluctuations always have small    
amplitudes ($|h| \ll 1$), we    
would then need to be in the regime of temperature and    
scales such that $|\beta V(r) | \ll1$, so that  
$h(r) \sim -\beta V(r)$.   
The potential should  
thus be attractive at smaller scales, as the system is  
positively correlated at those scales.  A $k^{-2}$ behaviour at 
large $k$, which is often 
used (see Figure \ref{fig3}) as an initial condition  
approximating cosmological models (CDM type) in 
this regime (beyond the ``turn-over''), may be obtained 
from an attractive $1/r$ potential, e.g.\  
\be   
V(r) = \frac{1}{r^2} -\eta \frac{e^{-\mu r}}{r} \, . 
\label{full-potential}
\ee By modifying the parameters $\mu$ and $\eta$, as well as the 
temperature, both the amplitude of the $P(k)$ and the location of 
a change 
from a  $P(k) \sim k^{-2}$ to a behaviour 
$P(k) \sim k$ for small $k$ can be controlled.
The potential (\ref{full-potential})  
is repulsive at short distances, but it may be necessary to make
it more strongly repulsive in order to to ensure the system is not
unstable to collapse \cite{ruelle}. Such issues, as well as the
practical numerical implementation of these observations will
be investigated in forthcoming work. Once this method has been 
implemented to set up the initial conditions, it will naturally 
be interesting to understand the effect of this change in initial 
conditions on  the dynamical evolution of gravitational N-body simulations.

\section {Acknowledgments} 
We thank Dominique Levesque for useful discussions about the OCP and
Charles Radin for information about pinwheel tilings.
FSL thanks the Swiss National Science Foundation    
for post-doctoral support. The work of JLL was   
supported by NSF grant DMR 98-13268 and AFOSR grant AF 49620-01-1-0154.

\end{document}